\documentclass[twocolumn,secnumarabic,amssymb,nobibnotes,aps,prl,reprint]{revtex4-1}
\usepackage[separate-uncertainty=true,multi-part-units=single]{siunitx}
\usepackage{graphicx}
\usepackage{bm}
\usepackage[usenames,dvipsnames]{color} 

\usepackage[version=3]{mhchem} 

\definecolor{mathematicaRed}{rgb}{0.92,0.386,0.209}
\definecolor{mathematicaBlue}{rgb}{0.37,0.51,0.71}
\definecolor{mathematicaGreen}{rgb}{0.56,0.69,0.19}

\definecolor{kcolor2}{rgb}{0.0196, 0.6, 0.6902}
\definecolor{kcolor3}{rgb}{0.643, 0.741, 0.0392}
\definecolor{kcolor4}{rgb}{1.0, 0.6509, 0.0823}
\definecolor{kcolor5}{rgb}{1.0, 0.1804, 0.0}

\definecolor{colorElectrophoresis}{rgb}{0.92,0.386,0.209}
\definecolor{colorPressure}{rgb}{0.37,0.51,0.71}
\definecolor{colorGravity}{rgb}{0.56,0.69,0.19}

\begin{document}


\title{Density dependent speed-up of particle transport in channels}

\author{Karolis Misiunas}
\email[correspondence: ]{karolis@misiunas.com}
\author{Ulrich F. Keyser}
\email[correspondence: ]{ufk20@cam.ac.uk}
\affiliation{Cavendish Laboratory, University of Cambridge, UK}
\date{\today}

\begin{abstract}
  Collective transport through channels shows surprising properties under  one-dimensional confinement: particles in a single-file exhibit sub-diffusive behaviour, while liquid confinement causes distance-independent correlations between the particles. Such interactions in channels are well-studied for passive Brownian motion but driven transport remains largely unexplored. Here, we demonstrate gating of transport due to a speed-up effect for actively driven particle transport through microfluidic channels. We prove that particle velocity increases with particle density in the channel due to hydrodynamic interactions under electrophoretic and gravitational forces. Numerical models demonstrate that the observed speed-up of transport originates from a hydrodynamic piston-like effect. Our discovery is fundamentally important for understanding protein channels, transport through porous materials, and for designing novel sensors and filters.
\end{abstract}

\maketitle


On the nano- and micron-scale, electric fields acting on ionic charges provide the dominant driving force for transport in aqueous environments. The so-called electrophoresis of charged particles is important for many fields including filtration and separation technologies. In these systems, it is commonly assumed that electrophoretic motion is independent of particle-particle distance and hence density. Any long-range interactions are routinely neglected in the thin Debye layer limit~\cite{Reed1976} because electroosmotic and Stokes contributions to the flow cancel each other in the bulk. 
However, this symmetry can be broken by the presence of a solid boundary, leading to significant particle interactions. The most extreme confinement of particles is found in channel geometries approaching the single-file limit. Examples include protein channels in biology~\cite{truskey2009transport}, nanopore sensors~\cite{Clarke2009,Bell2016a}, porous rocks, and filtration membranes~\cite{ingham1998transport}. In all these cases, channels accommodate the transport of charged ions, molecules, or particles. 
Particle interactions in channels have been extensively studied, especially in the free Brownian motion regime~\cite{Diamant2009,Wei2000,Cui2002,Misiunas2015}. In microfluidic systems, particles experience long-range hydrodynamic interactions that are independent of inter-particle distances~\cite{Misiunas2015}. These interactions arise due to momentum transfer from the moving particle to the liquid and then from the liquid to other particles~\cite{Lauga2009}. 
However, in the case of particle electrophoresis in channels, recent theoretical analyses found no long-range interactions~\cite{Hsu2005,Hsu2007}. On the contrary, the studies predict that the inter-particle interactions only extend to distances similar to the channel width. The difference between experimental~\cite{Misiunas2015} and theoretical studies~\cite{Hsu2005,Hsu2007} raises fundamental questions on the relevance of interactions for driven particle transport.

One striking example of the profound effects of particle-particle interactions on single-file transport is the stochastic gating observed in highly selective ion channels~\cite{hille2001ionic}. Such systems with diameters of a few angstroms allow the passage of ions via the multi-ion `knock-on' mechanism~\cite{Roux2017}. The stochastic transport leads to abrupt variations in the observed ionic currents in the time domain, known as gating~\cite{hille2001ionic}, often associated only with nanoscale systems. In this paper, we show that distance-independent hydrodynamic interactions present in microfluidic channels~\cite{Misiunas2015} gives rise to stochastic behaviour akin to gating in biological ion channels.

\begin{figure}[htb!]
  \centering
  \includegraphics[width=8.2cm]{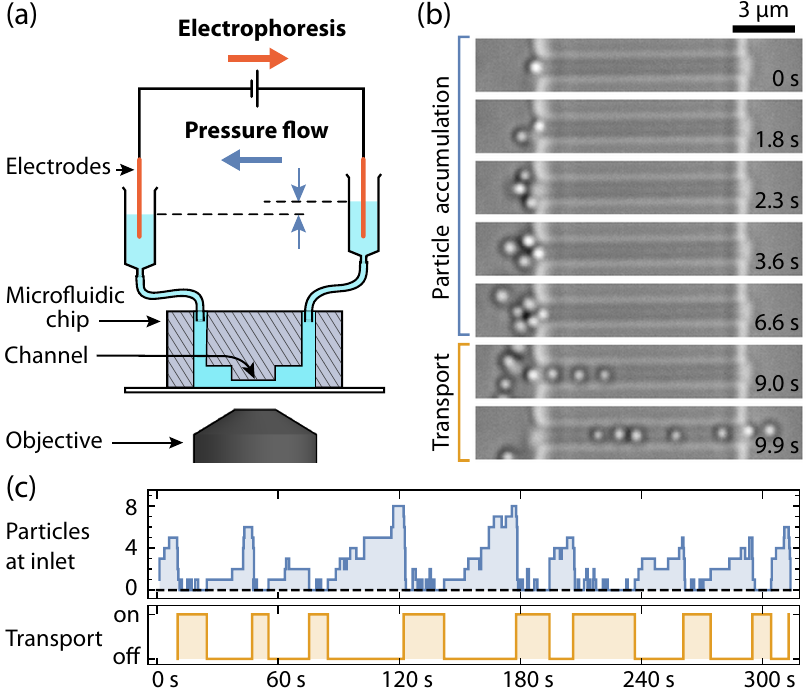}
  \caption{\label{dencity_filter}
  (a)~Schematic of experimental setup that allows for simultaneous control of the electric field and the pressure across the channel. 
  (b)~Micrographs show particle accumulation at the channel inlet with no transport because pressure flow dominates. However, at high particle densities, the particle-particle interactions dominate, thus allowing electrophoretic transport. 
  (c)~Number of particles at the channel inlet (blue) and transport (orange) shown as a function of time. Transport exhibits gating like behaviour between `on' and `off'. 
  } 
\end{figure}

In our experiments, we investigate the influence of interactions between driven particles inside microfluidic channels. Our setup allows us to directly quantify the role of particle-particle interactions during electrophoresis in single-file channels. We simultaneously control electric fields and pressure-driven flows, while also permitting direct imaging of particle motion. 
Figure~\ref{dencity_filter}a shows an illustration of our setup. Inside the microfluidic chip, two reservoirs are connected by narrow channels of length $L=\SI{10.0(5)}{\um}$ and rectangular cross-section $\SI{750(50)}{\nm} \times \SI{750(50)}{\nm}$. 
The chips are fabricated via replica moulding of polydimethylsiloxane (PDMS), where the master is made using e-beam lithography and photo-lithography; more details are published elsewhere~\cite{Pagliara2011,SupllementaryEnhanced}. A PDMS copy is air plasma bonded to a glass slide that is coated with a sub $\SI{100}{\um}$ PDMS layer, ensuring that all channel walls are made out of the same material~\cite{Deshpande2015}.

An assembled chip is filled with \ce{KCl} solution containing spherical polystyrene particles of diameter $2a = \SI{505(8)}{\nm}$ (Polysciences Inc.). Their motion is imaged through an inverted optical microscope with a high numerical aperture oil immersion objective ($100 \times$; NA~1.4; UPLSAPO) and recorded using a camera (Mikrotron MC1362) at a rate of 200 frames per second. Afterwards, the particle trajectories are extracted using established image analysis techniques~\cite{Dettmer2014}.

Particles are actively driven by an external electric field or/and a pressure flow. 
The electric field is applied using two Ag/AgCl-electrodes submerged in the external reservoirs. Electric potentials up to \SI{1}{\volt} are applied using a digital to analogue converter (NI-USB-6211) controlled by a computer. The pressure flow is controlled by adjusting the relative height of the external reservoirs. After assembling the chip, we find the pressure equilibrium by adjusting the pressure until particles stop migrating to either end of the channel.

Balancing pressure and electrophoretic forces in our microfluidic channels gives rise to a regime that resembles stochastic gating found in biological channels. 
Figure~\ref{dencity_filter}b shows two distinct states that we define as particle accumulation and transport. 
During the accumulation, particles are electrophoretically pulled towards the channel, but the opposing pressure flow inside the channel prevents transport across. As a result, the particle number increases over time at the left inlet~\cite{Rempfer2016,Hoogerheide2014}. 
Eventually, the transport starts when two or more particles randomly enter the channel. In this case, electrophoresis transiently dominates over the pressure and allows for transport. The transport ends when the last particle exits the channel.

Figure~\ref{dencity_filter}c shows the channel switching between the accumulation and the transport states. The blue line indicates that particles need to accumulate at the inlet, and that transport is possible only after 2 or more particles are present.
The resulting transport (orange lines) resembles stochastic gating found in biological ion channels. Importantly, in our system the channel's conformation is fixed and gating is a consequence of competition between different physical forces. 
In order to elucidate the origin of the gating effect we performed experiments investigating both pressure and electrophoretically driven transport separately.

\begin{figure}[htb!]
  \centering
  \includegraphics[width=8cm]{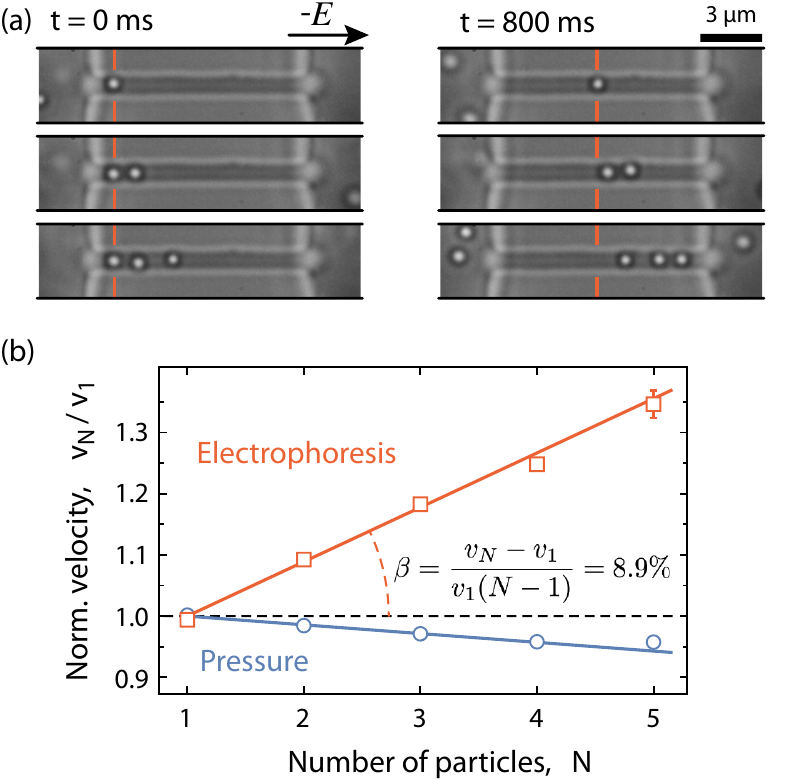}
  \caption{\label{vN_vs_N_electrophresis}
  Particle speed-up during electrophoresis. (a)~The left column shows micrographs with one, two, and three particles in the channel with leftmost particles aligned to the red line. The negatively charged particles migrate in an electric field for \SI{800}{\milli\second} and the resulting micrographs are shown on the right column. Red lines guide the eye. Three particles traveled further than 2 and 1 after \SI{800}{\milli\second}.(b)~Normalised particle velocity ($v_N/v_1$) as a function of particle number $N$  for electrophoretic transport (red) and for pressure-driven transport (blue). Errors are smaller than the marker sizes, except for $N=5$. Lines are weighted linear fits. The velocity linearly increases with $N$ under electophoresis but reduces under pressure flow. 
 }
\end{figure}

We start by measuring the velocity of 1, 2 and 3 particles in a channel driven only by electric fields, as shown in Figure~\ref{vN_vs_N_electrophresis}a. At $t=0$ the leftmost particles are aligned to the red line. After \SI{800}{\milli\second} the  snapshot shows that the two particles travelled further than one particle; and three made it further than two, illustrating the unexpected increase in velocity with particle number.

Quantitatively, we analyse the particle trajectories by measuring the velocity as a function of particle number, $N$, inside the channel. Only particles separated by more than \SI{1.2}{\um} are analysed, thus excluding any close range effects~\cite{Misiunas2015,Hsu2005}. In addition, we disregard all parts of the trajectories within \SI{0.5}{\um} of the channel ends to ensure diffusion coefficients are constant~\cite{Dettmer2014a}. The remaining trajectory segments are averaged while retaining the particle count.

\begin{figure*}[tb!]
  \centering
  \includegraphics[width=16.4cm]{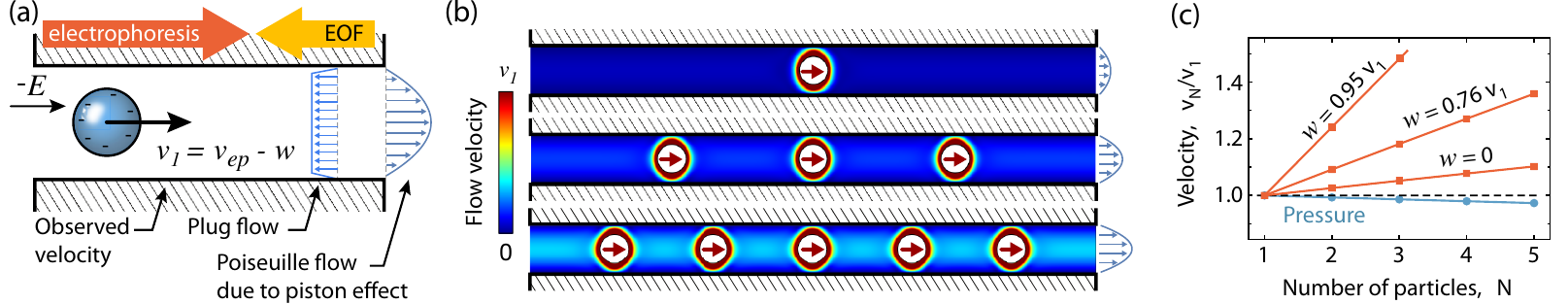}
  \caption{\label{simulation}
  Simulation results.
  (a) The applied external field, $E$, drives a negatively charged particle through a finite channel at velocity $v_{ep}$, giving rise to a Poiseuille flow due to a piston effect. Simultaneously, negatively charged walls induce an electro-osmotic plug flow (EOF) in the opposite direction at velocity $w$.
  (b) Simulated flow velocities for one, three, and five particles driven through a channel by an electric field (red arrow indicates the direction of force) for $w=0$. The Poiseuille flow increases with particle number in the channel. 
  (c) The predicted normalised particle velocity, $v_N/v_1$, increases with $N$ for particles driven by electric field (red), but decreases for pressure driven particles (blue).
  }
\end{figure*}

The red points in Figure~\ref{vN_vs_N_electrophresis}b show normalised velocity ($v_N/v_1$) as a function of $N$ for electrophoretic motion, while keeping the pressure flow at zero (estimated from $\sim{}38000$ video frames of more than $650$ particles). The normalisation velocity is $v_1 = \SI{17.48(8)}{\um/\second}$ at an applied potential of \SI{200}{\milli\volt}. The velocity linearly increases with $N$, clearly contradicting the previous theoretical predictions~\cite{Hsu2005,Hsu2007}.

After driving the particles by electric fields we now investigate the velocity under pressure driven flows. The blue data in Figure~\ref{vN_vs_N_electrophresis}b shows the results performed in the same microfluidic systems with particles from the same batch. The normalisation velocity is $v_1 = \SI{68.10(6)}{\um/\second}$ at height difference of \SI{11.9}{\milli\metre} that produces the pressure flow. The velocity decreases slightly with the number of particles in contrast to the electrophoresis results.

The stark difference between the particle interactions, allows us to explain the observed gating behaviour in Figure~\ref{dencity_filter}b. The pressure-driven velocity stays roughly constant for all $N$ and blocks transport for $N=1$. However, for $N \ge 2$, electrophoretic velocity increases due to the speed-up (Figure~\ref{vN_vs_N_electrophresis}b) and overcomes the opposing pressure flow. Thus when more than two particles enter the channel, electrophoresis transport is permitted, until $N=0$ and accumulation starts again.

Having explained the observed gating, we now quantify the underlying phenomena.
To compare the different driving forces we define an \emph{interaction coefficient}: $\beta \equiv \Delta v / v_1$, where $\Delta v=v_{N+1}-v_N$ is the averaged speed-up due to each additional particle. $\beta$ can be extracted from the slope of the curves in Figure~\ref{vN_vs_N_electrophresis}b. For electrophoresis, $\beta = \SI{8.9(2)}{\%}$ at \SI{2}{mM} \ce{KCl} and pH~7.2. Meanwhile, for the pressure driven transport $\beta = \SI{-1.43(07)}{\%}$. The interaction coefficient corresponds to the fraction of velocity gained with each additional particle inside the channel. 


As we are unaware of any analytical solutions for electrokinetic transport, we model the particle velocity using a numerical model implemented using COMSOL Multiphysics (v4.4). Our model is accessible online~\cite{MisiunasComsolCode2017}. Before we discuss the details of the model we introduce the relevant electrokinetic effects in Figure~\ref{simulation}a. We use an axial-symmetric channel containing a spherical particle in a uniform electric field ($E$), as shown in Figure~\ref{simulation}a. The particle carries a negative surface charge, while the charge on the channel walls has the same sign but varied magnitude. We assume a low Reynolds number regime and that flows have no-slip boundary conditions at the walls.
The electrophoretic force drives the particle through the channel at a velocity $v_{ep}$ resulting in the liquid being pushed forwards, resembling a piston effect. The resulting Poiseuille flow is indicated on the right of Figure~\ref{simulation}a. In the case where the channel walls carry a negative charge, the well-known electro-osmotic plug flow (EOF) develops with velocity $w$  moving in the opposite direction to $v_{ep}$~\cite{Kirby2004a}.

In contrast to the literature~\cite{Hsu2005,Hsu2007}, we adopt open boundary conditions\footnote{Alternatively, the periodic boundary conditions can be used with equivalent results. See the Supplementary Information for details.} at the inlets to account for the finite channels, which allow for critically important flows, thus enabling hydrodynamic interactions~\cite{Misiunas2015}.

Figure~\ref{simulation}b summarises the key results of our simulations with color maps depicting fluid flows for $N=1$, 3, and 5. The first row of Figure~\ref{simulation}b shows the flow velocity for N=1 due to electrophoretic body force (indicated by the red arrow in the particle). The particle's motion induces a finite Poiseuille flow as illustrated by the profile on the right. Increasing $N$ to 3 and 5 in rows two and three, receptively, increases the magnitude of the Poiseuille flow. The enhanced net flow throughout the channel is easily observed by the change in color from dark to light blue. Importantly, increased Poiseuille flows increase the velocity of all particles.

Figure~\ref{simulation}c shows quantitative predictions of our simulations for particles driven by electrophoresis (red) and pressure driven flows (blue). The simulation parameters were selected to match our experiments with $2a = \SI{500}{\nm}$; $L = \SI{10}{\um}$; and $2R= \SI{840}{\nm}$, where this diameter corresponds to the channel's cross-section area in the experiments~\cite{SupllementaryEnhanced,Pagliara2014a}. The first observation is that the slope of $v_N/v_1$ clearly depends on the type of driving force; similar to the experiments. 
For particles driven by electrophoresis $v_N/v_1$ linearly increases with $N$ (red lines). The corresponding $\beta = \SI{2.6}{\%}$ for simulation with no EOF ($w=0$). 
In contrast, particles carried by a pressure flow exhibit a decreasing $v_N/v_1$ with $N$ (blue curve) and a negative $\beta=\SI{-0.69}{\%}$. This decreasing velocity is due to particles perturbing the optimal flow profile, thus slowing down the pressure induced flow that carries them~\cite{Happel1973}.

The simulations also reveal that $\beta$ depends on $w$. Figure~\ref{simulation}c shows simulations for three different surface charge densities that correspond to $w/v_{ep}=0.0,\; 0.76$, and $0.95$. The resulting $\beta$ coefficients are \SI{2.6}{\%}, \SI{8.9}{\%}, and \SI{24.1}{\%}, showing that EOF velocity increases $\beta$.

The dependence of $\beta$ on $w$ can be explained by considering a linear superposition of the flows induced by electrophoresis and EOF. For one particle, $v_1=v_{ep} - w$, as illustrated in Figure~\ref{simulation}a. For $N>1$, we can approximate $v_N \approx v_{ep,N} - w$, where $v_{ep,N}$ indicates the electrophoretic velocity for $N$ particles without the EOF and the approximation comes from non-linear components that are small and neglected for our salt concentrations~\cite{Liu2014}.
As a result, the EOF contributions cancel, making $\Delta v$ independent of the magnitude of EOF and thus $\beta \approx \Delta v / (v_{ep} - w)$. The $\beta$ is a hyperbolic function with respect to $w$ and $v_1$. We can maximise $\beta$ with electrophoretic velocity matched by the EOF velocity and can minimize the interaction coefficient when $w=0$ (or negative). One important conclusion is that $\beta$ is controlled by the channel surface charge and hence the EOF.


\begin{figure}[ht!]
  \centering
  \includegraphics[width=\linewidth]{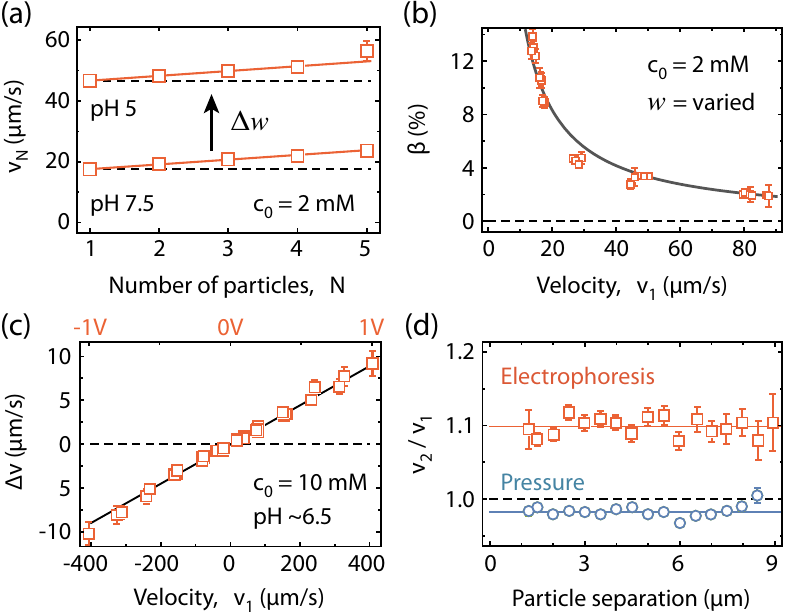}
  \caption{\label{electrophoresis_experiments}
  (a) Velocity as a function of $N$ for pH~5 and pH~7.5, which changes the surface charge density on the walls, thereby changing $w$.
  (b) Interaction coefficient, $\beta$, is shown as a function of particle velocity $v_1$. 
  (c) Speed-up velocity, $\Delta v$, as a function of $v_1$ and the applied voltage (top axis).
  (d) Normalised two particle velocity, $v_2/v_1$, is independent of the inter-particle distance.
  }
\end{figure}

In experiments, we are able to explore the relationship between $\beta$ and $w$ by adjusting the EOF. Figure~\ref{electrophoresis_experiments}a shows two measurements at pH~5 and pH~7.5 as a function of $N$. Decreasing pH decreases surface charge on the PDMS~\cite{Kirby2004a} and thus reduces $w$. $v_1$ increases from \SI{17.5(1)}{\um\per\second} to \SI{46.7(1)}{\um\per\second}, while $\Delta v$ stays roughly the same at $\Delta v = \SI{1.56(7)}{\um\per\second}$ and $\Delta v = \SI{1.58(6)}{\um\per\second}$. As a result, the interaction coefficient decreases from $\beta = \SI{8.8(3)}{\%}$ to $\beta = \SI{3.4(2)}{\%}$, proving the relationship with $w$.

Figure~\ref{electrophoresis_experiments}b shows $\beta$ as a function of $v_1$ for a wide range of $w$ values. $w$ was changed by varying the pH in the range from 5 to 10, and by exposing the channels to water for 1~hour, 12~hours, or 24~hours, which reduces the surface charge on the PDMS~\cite{Bodas2007}. The measurements follow a hyperbolic curve, as expected. A weighted fit (grey line) to $\beta=\Delta v / v_1$ gives $\Delta v = \SI{1.59}{\um\per\second}$. The rightmost data points in Figure~\ref{electrophoresis_experiments}b have the smallest $w$ that correspond to $\beta = \SI{1.80(14)}{\%}$. This value is close to the simulation prediction for $w=0$. 
Since we do not measure $w$ directly in our experiment, it is the only fitting parameter in our model. The simulations agree very well with the experimental data, suggesting that our model captures the important physics.

Figure~\ref{electrophoresis_experiments}c shows the speed-up as a function of applied potential. $\Delta v$ linearly increases with the applied electric potential (and $v_1$), while $\beta$ stays constant at \SI{2.3}{\%}. $\Delta v$ and hence the interaction strength scales linearly with the driving potential, thus affirming our choice the relative quantity -- $\beta$.

In addition, we provide further evidence that the particle speed-up is indeed due to distance-independent hydrodynamic interactions. Figure~\ref{electrophoresis_experiments}d shows normalised two particle velocity, $v_2/v_1$, as a function of their septation distance. Evidently, $v_2/v_1$ is independent of the separation distance for electrophoresis. We see the same characteristic in our hydrodynamics simulations (Figure~S1) and similar distance-independent interactions were observed before but between freely diffusing particles in channels~\cite{Misiunas2015}. Combining all the experimental evidence with the simulations, suggests that the increase of particle velocity with $N$ is caused by distance-independent hydrodynamic interactions.

\begin{figure}[ht!]
  \centering
  \includegraphics[width=8cm]{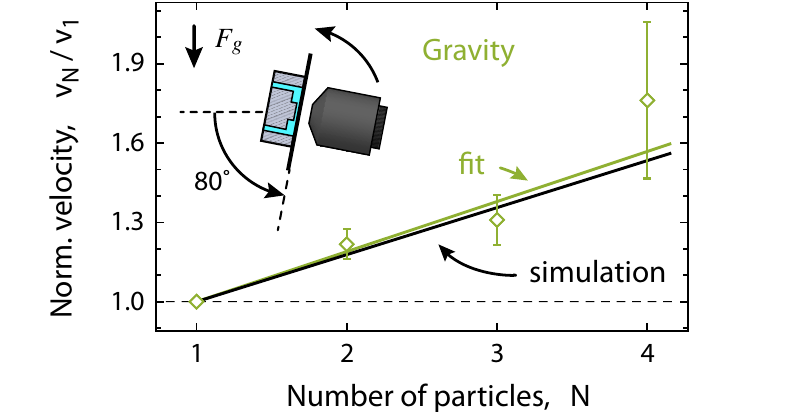}
  \caption{\label{velocity_vs_N}
  Normalised particle velocity as a function of the particle number inside a channel for gravity propelled particles (green) compared to simulation results (black). 
  Inset illustrates the setup for gravity experiments.
  }
\end{figure}

Finally, we show that $v_N$ increases for other body forces by changing the driving force to gravity. Gravity experiments are performed in the same microfluidic channels, but with gold particles of diameter $2a=\SI{400(50)}{\nm}$ (supplied by CytoDiagnostics Inc.; in \SI{0.1}{\milli M} PBS and \SI{1}{\milli M} \ce{KCl}). Gold particles have much higher mass density than water and thus the gravitational force is larger than for the polystyrene system used before. An assembled chip is mounted on a custom built rotatable microscope, as shown in Figure~\ref{velocity_vs_N} inset. In contrast to the experiments using the pressure or the electric fields, gravity pulls the particles directly downwards without focusing them into the channels. This reduces the number of particles entering the channels and thus increases the uncertainty in our measurements ($\sim\,313000$ frames recorded at 30~fps of more than 330 particles). In addition, the net force on the particles reaches only about \SI{6}{\femto\newton} (corresponding to $v_1 = \SI{0.23(2)}{\um/\second}$). Compared to thermal energy of $\sim \SI{4}{\pico\newton\nano\meter}$, fluctuation due to Brownian motion are significant and further increase variability.  

Despite the experimental challenges, Figure~\ref{velocity_vs_N} shows that gravity also exhibits the speed-up effect. A weighted fit gives an interaction coefficient of $\beta = \SI{18.9(34)}{\%}$, which agrees well with the corresponding simulation value of  
\SI{17.9}{\%}. The simulation parameters were set to experimental values $2a=\SI{400}{\nm}$; $2R= \SI{840}{\nm}$ and had no fitting parameters~\cite{SupllementaryEnhanced}. Our experiment confirms that the speed-up effect is universal for actively driven particles that are propelled by a body force.


The speed-up effect has important implications for understanding natural phenomena. 
We have demonstrated that it can cause gating like behaviour without conformational changes of the channel. This raises fundamental questions on how protein channels operate. 
In addition, the presence of constructive interactions suggests that particle transport rate through channels is a super-linear function of particle density. As a result, membranes with embedded channels should permit faster transport at higher particle densities. This can explain some of the peak tailing observed in electrochromatography or chromatography centrifuges because collectively particles travel faster than the isolated particles~\cite{Sun2007}.

Although we only explored single-file transport, the speed-up is not limited to tightly confined particles and should persist for larger channels ($a/R < 0.3$). In a large channel, each particle's piston-like contribution is small but the collective action of many particles should result in a net speed-up. Thus the speed-up effect might affect a wide range of confined systems, including self-propelled particles in confinement~\cite{Wioland2016a}.

The phenomenon also enables novel technological applications. In particular, the particle accumulation at the inlet can be utilised to control a chemical reaction rate for manufacturing dimer particles. Or it could be used for particle separation. However, further investigations are necessary to fully realise all the technological opportunities presented by our discovery.

In conclusion,
we have shown that particle transport velocity through channels increases with the particle number. We experimentally demonstrated that this happens with electrophoretically and gravity propelled particles but does not happen for pressure propelled particles. Our models suggest that the interactions are carried by hydrodynamics, where a piston-like motion of particles induces a flow throughout the entire channel. These findings have far-reaching implications for transport through protein channels and enable novel technological applications.

\begin{acknowledgments}
  We are grateful to Soichiro Tottori, Stefano Pagliara, Eric Lauga, Nicholas A\@.W\@.~Bell, Vahe Tshitoyan, Jehangir Cama, and Alexander Ohmann for useful discussions.
  K\@.M\@.~and U\@.F\@.K\@.~acknowledge funding from an ERC consolidator grant (Designerpores 647144). 
\end{acknowledgments}

\nocite{Papanastasiou1992}

\bibliographystyle{naturemag}
\bibliography{bib_library}

\end{document}